\documentclass[aps, prb,showpacs,twocolumn, floatfix]{revtex4}
\usepackage{amsmath}
\usepackage{amssymb}
\usepackage{mathtools}
\usepackage{epsfig}
\usepackage{graphics}
\usepackage{color}
\usepackage{bm}
\usepackage{amssymb}
\usepackage{graphicx}
\usepackage{epsfig}

\begin{document}

\preprint{Sepper-Lebed-PRB}

\title{Possible restoration of superconductivity in the quasi-one-dimensional
conductor Li$_{0.9}$Mo$_6$O$_{17}$ in feasibly high pulsed magnetic
fields, $H \simeq 100 \ T$.}

\author{O. Sepper}
\author{A.G. Lebed$^*$}

\affiliation{Department of Physics, University of Arizona, 1118 E.
4-th Street, Tucson, AZ 85721, USA}

\begin{abstract}
We present a theoretical study of restoration of superconductivity
in the form of the triplet reentrant superconducting phase in a
quasi-one-dimensional (Q1D) conductor. Substitution of known band
and superconducting parameters of the presumably triplet Q1D
superconductor Li$_{0.9}$Mo$_6$O$_{17}$ into our theoretical
equations shows that such restoration can happen in feasibly high
non-destructive pulsed magnetic field of the order of $H \simeq
100 \ T$. We investigate in detail how small inclinations of a
direction of a magnetic field from its best experimental geometry
decrease superconducting transition temperature of the reentrant
phase, which is important for its possible experimental discovery.
\end{abstract}


\maketitle

\section{Introduction}

Physical properties of quasi-one-dimensional (Q1D) superconductors
have been intensively studied since the discovery of
superconductivity in the Bechgaard salts - superconductors in the
chemical family (TMTSF)$_2$X, where X= ClO$_4$, PF$_6$, etc [1,2].
Early experiments alluded to the unconventional nature of these
superconductors, as non-magnetic impurities destroyed
superconductivity [3,4] and the so-called Hebel-Slichter peak in
the NMR data was not observed [5]. Recent experiments [6] have
firmly supported these early findings. Despite 30 years of
intensive investigations, the nature of superconductivity in the
Bechgaard salts is still controversial. On the one hand it was
shown [7-9] that superconducting phase in (TMTSF)$_2$ClO$_4$
compound is more likely of d-wave type with zeros on Q1D Fermi
surface. On the other hand, in (TMTSF)$_2$PF$_6$ compound there is
still some chance of a spin triplet superconducting pairing
[10-12].

Triplet superconductors with layered Q1D Fermi surfaces (FS) were
theoretically shown to exhibit very unusual magnetic properties in
a high magnetic field perpendicular to the chains and parallel to
the layers. More precisely, it was demonstrated [13-16] that
superconductivity can be restored as a pure two-dimensional phase
when the magnetic field is high enough to localize conducting
electrons on Q2D layers. This happens when the typical "sizes" of
electron trajectories become less than the inter-plane distance
due to the so-called quantum $3D \rightarrow 2D$ dimensional
crossover [2]. (Note that this phenomenon, which we call reentrant
superconductivity, was first suggested in Ref.[13] and is
different from the quantum limit (QL) superconductivity, suggested
for a pure 3D case by Tesanovic and Razolt [17] and for a Q1D case
by us [18].)

In this context, the possible triplet superconductor
Li$_{0.9}$Mo$_6$O$_{17}$ presents unique opportunity to study
superconductivity in high magnetic fields.
Li$_{0.9}$Mo$_6$O$_{17}$ is a Q1D, layered transition metal oxide,
structurally similar to the Bechgaard salts. It is metallic at
high temperatures and undergoes a superconducting transition at
$T_c\approx 2.2 \ K$. Recently, Mercure et al. [19] have shown
that superconductivity in the Li$_{0.9}$Mo$_6$O$_{17}$ compound
exceeds the so-called Clogston-Chandrasekhar paramagnetic limit
[20] by 5 times. Our theoretical analysis [21,22] of the
experimental curves in Ref. [19] showed excellent quantitative
agreement between the experimental properties and the theoretical
predictions for a spin-triplet superconductor in the absence of
Pauli destructive effects. Thus, the presumably triplet
superconductor Li$_{0.9}$Mo$_6$O$_{17}$ is a major candidate for
experimental discovery of reentrant superconductivity. We note
that the possibility of existence of reentrant superconductivity
phenomenon is restricted by such triplet phases where the
spin-splitting Pauli effects against superconductivity are absent.
The band and superconducting parameters of this compound have been
calculated in Ref. [21], along with the most likely triplet
pairing scenario presented in Refs. [21,22]. The latter shows that
a nodeless scenario of triplet superconductivity parametrized by
an order parameter that changes its sign on the two sheets of the
Q1D Fermi surface leads to the experimentally observed destruction
of superconductivity by pure orbital effects in a magnetic field
parallel to conducting chains.

The goal of this paper is twofold. First, we suggest experimental
discovery of the reentrant superconducting phase in the possible
spin-triplet, Q1D layered superconductor Li$_{0.9}$Mo$_6$O$_{17}$
using the currently experimentally available pulsed
(non-destructive) magnetic fields of order $H \simeq 100$ Tesla.
Our calculations show that such fields, parallel to the layers and
perpendicular to the conducting chains lead to the appearance of
the reentrant superconductivity phenomenon in
Li$_{0.9}$Mo$_6$O$_{17}$ below a reentrant transition temperature
of $T^*(H=100 T) \simeq T_c/2 \simeq 1$K. Second, we calculate the
angular dependence of $T^* (\alpha, H)$ to account for inclination
of the same magnetic field from the optimal geometry of the
corresponding experiment. The latter will allow to conduct the
experiments with necessary accuracy.

\begin{figure}
\centering
\includegraphics[width=93mm, height=65mm]{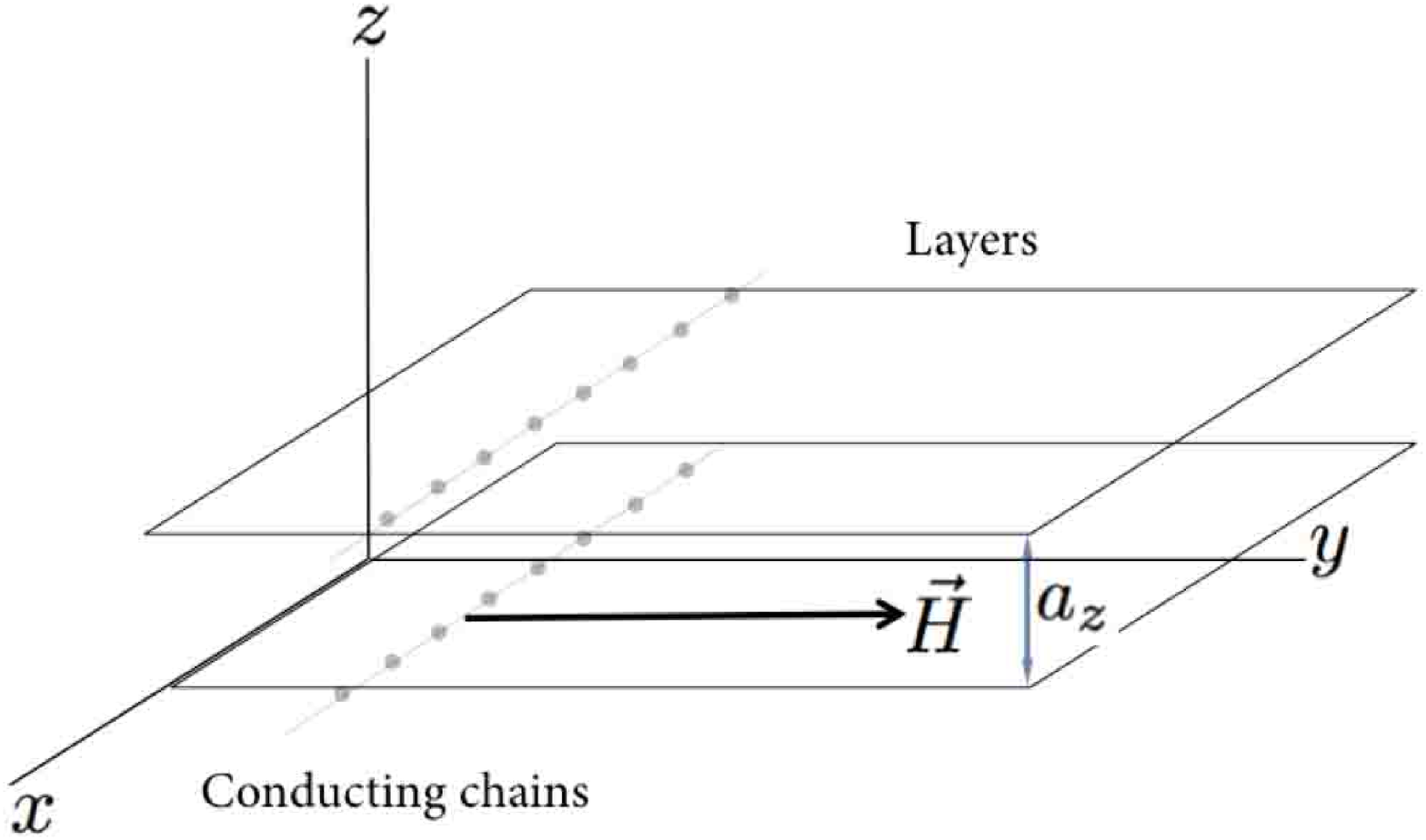}
\caption{The setup for observation of reentrant superconductivity
phenomenon in a layered Q1D compound with spin-triplet pairing.
Magnetic field is parallel to the layers and perpendicular to the
conducting chains.}
\end{figure}

\begin{figure}
\centering
\includegraphics[width=70mm, height=95mm]{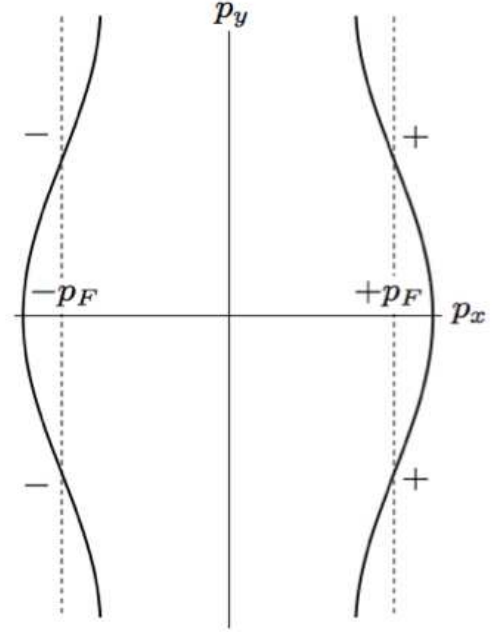}
\caption{Q1D Fermi surface consisting of two open, slightly
corrugated sheets extending in the $z$-direction, and centered at
the Fermi momentum $p_x=\pm p_F$. The spin-triplet superconducting
order parameter changes its sing on the two sheets of the Fermi
surface.}
\end{figure}

\section{Reentrant superconducting phase in Li$_{0.9}$Mo$_6$O$_{17}$}

For our present theoretical analysis of the reentrant phase, we initially consider a Q1D
conductor in a magnetic field parallel to the layers and perpendicular to the conducting
chains, as shown in Fig.1. The Q1D Fermi surface, shown in Fig.2, consists of two open,
slightly warped sheets centered at $p_x=\pm p_F$. The anisotropic electron spectrum,
with $t_x>>t_y>>t_z$ can be linearized on the right (+) and left (-) sheets as follows:

\begin{equation}
\epsilon^{\pm}(\vec{p}) = \pm v_F (p_x - p_F) - 2t_y\cos\left(p_ya_y\right) - 2t_z\cos\left(p_za_z\right) .
\end{equation}

\noindent The magnetic field and the vector potential have the following components:

\begin{equation}
\vec{H} = (0,H,0), \ \ \ \ \vec{A} = (0,0,-Hx).
\end{equation}

\noindent The Hamiltonian is obtained from Eq.(1) by the Peierls' substitution method [2,23]:

\[ (p_x - p_F) \rightarrow -i \frac{\mathrm{d}}{\mathrm{d} x}, \ \ \ \ p_ia_i \rightarrow p_ia_i - \frac{ea_i}{c}A_i, \]

\noindent Motion parallel to the magnetic field (along the $y$ direction) does not lead to twisting
of electron orbits, allowing us to consider the following reduced Hamiltonian for a system in a magnetic field:

\begin{equation}
\mathcal{\hat H}^{\pm} = \mp i v_F \frac{\mathrm{d}}{\mathrm{d} x} - 2t_z\cos\left(p_za_z + \frac{\omega_z}{v_F}x\right),
\end{equation}

\noindent where the transverse oscillation frequency is

\begin{equation}
\omega_z = \frac{ev_Fa_zH}{c}.
\end{equation}

\noindent The Schr\"{o}dinger-like equation, $\mathcal{\hat H}
\Psi_{\epsilon} = \epsilon\Psi_{\epsilon}$, for the Q1D wave
functions becomes

\begin{equation}
\biggl[\mp i v_F \frac{\mathrm{d}}{\mathrm{d} x}
-2t_z\cos\left(p_za_z +
\frac{\omega_z}{v_F}x\right)\biggr]\Psi^{\pm}_{\epsilon}(x;p_z) =
\epsilon \Psi^{\pm}_{\epsilon}(x;p_z),
\end{equation}

\noindent where the energy $\epsilon$ is measured with respect to
the Fermi energy. By direct substitution into Eq.(5) we can check
that the exact solutions for the Q1D electron wave functions are

\begin{multline}
{\Psi_{\epsilon}^{\pm}(x;p_z) = \frac{1}{\sqrt{2\pi v_F}}
\exp\left(\pm i \frac{\epsilon}{v_F}x\right) \times }\\
\exp\left\{  \mp  \frac{2it_z}{\omega_z}
\left[  \sin\left(   p_za_z+\frac{\omega_z}{v_F}x \right)
- \sin(p_za_z)\right] \right\}.
\end{multline}

\noindent The normalization of these wavefunctions is such that

\[\int_{-\infty}^{\infty} \Psi^{\pm}_{\epsilon}(x;p_z)
\Psi^{\pm}_{\epsilon'}(x;p_z)^*\, \mathrm{d}x = \delta(\epsilon-\epsilon').\]

Having obtained the wave functions, we construct the finite temperature
Green's functions according to the standard procedure [24]:

\begin{equation}
G^{\pm}_{i\omega_n}(x,x';p_z) = \sum_{\epsilon}\frac{\Psi^{\pm}_{\epsilon}(x;p_z)
\Psi^{\pm}_{\epsilon}(x';p_z)^*}{i\omega_n-\epsilon},
\end{equation}

\noindent where the Matsubara frequency for fermions is $\omega_n
= \pi T (2n+1)$. Performing the sum in Eq.(7) we obtain the expressions for
the Green's functions:

\begin{equation}
G^{\pm}_{i\omega_n}(x,x';p_z) = -\frac{i}{v_F}sgn(\omega_n) e^{\mp\omega_n(x-x')/v_F}
g^{\pm}(x,x';p_z),
\end{equation}

\noindent where

\begin{multline}
g^{\pm}(x,x';p_z) = \\
\exp\left\{  \mp\frac{2it_z}{\omega_z} \left[   \sin\left(p_za_z+\frac{\omega_z}{v_F}x\right)
- \sin\left(p_za_z+\frac{\omega_z}{v_F}x'\right)\right]\right\}
\end{multline}

The equation to determine superconducting order parameter is
obtained from Gor'kov's equations for non-uniform
superconductivity [24]. To this end, we take the simplest spin
triplet order parameter that changes its sign on the two sheets of
the Q1D Fermi surface. Note that it satisfies experimental data
[19] and is insensitive to Pauli spin-splitting effects,

\begin{equation}
\hat{\Delta} = \hat{I}sgn(p_x)\Delta(x),
\end{equation}

\noindent where $\hat{I}$ is a unit matrix in spin space, and $sgn(\pm p_F) = \pm 1$,
while $\Delta(x)$ is the solution of the so-called gap
integral equation

\begin{equation}
\Delta(x) = g\int K(x,x')\Delta(x')\mathrm{d}x',
\end{equation}

\noindent where $g$ is the effective electron coupling constant,
and the kernel is

\begin{equation}
K(x,x') = T\sum_{\omega_n} \biggl< G^{+}_{i\omega_n}(x,x';p_z)
G^{-}_{-i\omega_n}(x,x';-p_z)\biggr>_{p_z} .
\end{equation}

\noindent The products of Green's functions in Eq.(12) are averaged
over the momentum component $p_z$. By using the expressions
for the Green's functions in Eq.(9) it can be shown that

\begin{multline}
{ \biggl< g^{+}(x,x';p_z)g^{-}(x,x';-p_z)\biggr>_{p_z}  = }\\
{\biggl< \exp\left\{  \frac{8it_z}{\omega_z}\sin(p_za_z)
\sin\left[\frac{\omega_z}{2v_F}(x-x')
\right]\sin\left[\frac{\omega_z}{2v_F}(x+x')\right] \right\} \biggr>_{p_z}} \\
=J_0\left\{  \frac{8it_z}{\omega_z}\sin\left[\frac{\omega_z}{2v_F}(x-x')
\right]\sin\left[\frac{\omega_z}{2v_F}(x+x')\right] \right\},
\end{multline}

\noindent where we have used the Bessel function definition

\[\biggl< e^{\pm iy\sin(t)}\biggr>_{t} =
\frac{1}{2\pi}\int_{-\pi}^{\pi}e^{\pm iy\sin(t)}\mathrm{d}t = J_0(y).\]

\noindent The summation over both the positive and the negative Matsubara
frequencies in Eq.(12) results in a factor

\begin{equation}
\sum_{\omega_n} e^{-2\omega_n(x-x')/v_F}  =
\frac{1}{2\sinh\left[ \frac{2\pi T}{v_F}|x-x'|\right]}.
\end{equation}

\noindent Substituting the results of Eqs.(13),(14) into the definition of
the kernel in Eq.(12), we obtain the gap integral equation
for the superconducting order parameter [13]:

\begin{multline}
\Delta(x) = \frac{g}{2} \int_{|x-x'|>d} \frac{2\pi T \Delta(x')
\mathrm{d}x'}{v_F\sinh\left[\frac{2\pi T}{v_F}|x-x'|\right]} \times \\
J_0\left\{  \frac{8t_z}{\omega_z} \sin\left[\frac{\omega_z}{2v_F}(x-x')\right]
\sin\left[\frac{\omega_z}{2v_F}(x+x')\right]\right\},
\end{multline}

\noindent where $d\sim v_F/\Omega$ is the cutoff distance. Note
that the equation obtained above is quite general, and provides
various descriptions of superconductivity for Q1D layered
compounds. For example, it can be shown that for temperatures
$T\approx T_c$ and low enough magnetic fields Eq.(15) reproduces
the Ginzburg-Landau relations. In the quasi-classical regime of
high enough temperatures and low magnetic fields, Eq.(15)
reproduced both the descriptions of anisotropic 3D
superconductivity and the Lawrence-Doniach model.

Our goal is to study theoretically the solutions of Eq.(15) under
high magnetic fields that lead to the phenomenon of reentrant
superconductivity [13-16]. Using the superconducting and band
parameters for Li$_{0.9}$Mo$_6$O$_{17}$, we can show that the
reentrant phase can be achieved at fields of $H \simeq 100$ Tesla
at the transition temperature $T^*(H=100 \ T) \approx 1\
$K$\approx T_c/2$. To this end, we use the expansion of the Bessel
function in Eq.(15) with $8t_z/\omega_z$ as the small quantum
parameter:

\begin{equation}
J_0(\cdot\cdot\cdot) \approx 1 - \frac{4t_z^2}{\omega_z^2}
\biggl[1 - \cos\left(\frac{\omega_z}{v_F}(x-x')\right)\biggr],
\end{equation}

\noindent where in the subsequent approximation the Bessel
function expansion was averaged over $x+x'$ variables. We use the
defining relation for the zero-field critical temperature,

\begin{equation}
\frac{1}{g} = \int_{d}^{\infty}
\frac{2\pi T_c\mathrm{d}z}{v_F\sinh\left(\frac{2\pi T_c}{v_F}z\right)},
\end{equation}

\noindent and the following standard approximation:

\begin{equation}
\int_0^{\infty} \frac{1-\cos(\beta x)}{\sinh(x)}
\mathrm{d}x \approx \ln(2\beta \gamma) ,
\end{equation}

\noindent where $\gamma\approx 1.781$ is the exponential of the
Euler constant. Substituting the relations in Eqs.(16)-(18) into
the gap integral equation, and using the special trial solution
$\Delta(x)=\mathrm{const}.$, we obtain an analytical equation that
implicitly determined the critical reentrance temperature,
$T^*(H)$, as a function of magnetic field:

\begin{equation}
\ln\left[\frac{T_c}{T^*(H)}\right]= \frac{4t_z^2}{\omega^2_z(H)}
\ln\left[\frac{\gamma \omega_z(H)}{\pi T^*(H)}\right].
\end{equation}

\noindent For the solution (19) of the above equation, the
numerical values $t_z=14$K and $\omega_z(H) = ev_Fa_zH/c = 0.58H$
K/Tesla are obtained using the parameters for
Li$_{0.9}$Mo$_6$O$_{17}$ calculated in Ref. [21]. Thus, Eq.(19)
can be numerically solved to obtain the field dependence of the
reentrance transition temperature, shown in Fig.3. [Note that, in
Fig.3, the superconducting transition temperature is shown only
for high magnetic fields, $H \geq 90 \ T$, where the approximation
(16) and, thus, Eq.(19) are valid. Nevertheless, at lower magnetic
fields, superconductivity is still stable, although may have
exponentially small transition temperature (see Refs.[13-16]).]

\begin{figure}[h]
\centering
\includegraphics[width=90mm, height=70mm]{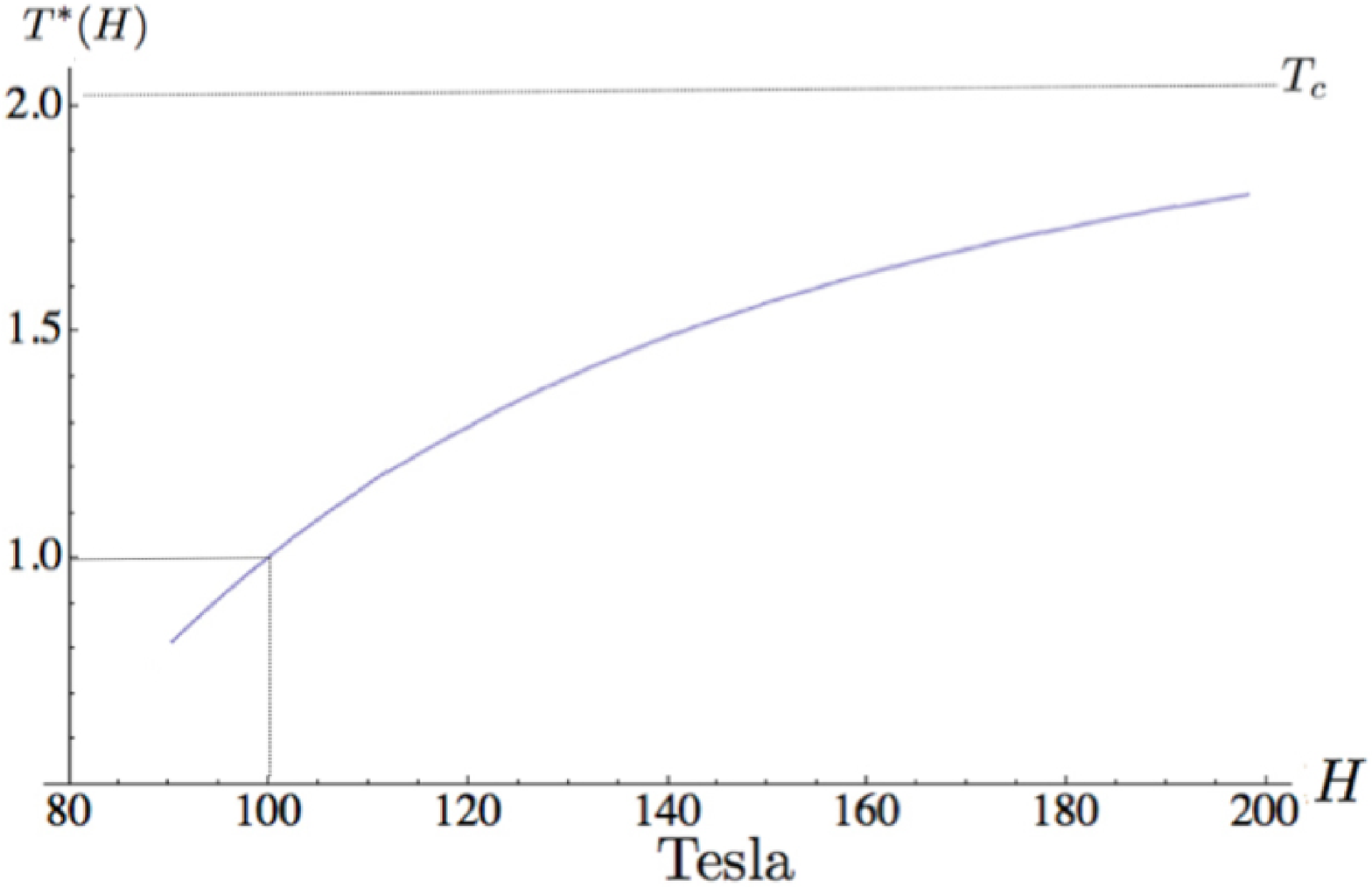}
\caption{Reentrant superconducting transition temperature (in
degrees Kelvin) for Li$_{0.9}$Mo$_6$O$_{17}$ as a function of
magnetic field oriented perpendicular to conducting chains and
parallel to the layers, as depicted in Fig.1.}
\end{figure}

\noindent We recall that the superconducting transition
temperature for Li$_{0.9}$Mo$_6$O$_{17}$ in the absence of
magnetic field is $T_c=2.2 \ K$. Based on the above curve, we can
see that the reentrant superconducting phase in
Li$_{0.9}$Mo$_6$O$_{17}$ can be obtained at the experimentally
available non-destructive pulsed magnetic field and temperature of

\[T^* \approx 1\mathrm{K}  \ \ \ \mathrm{at} \ \  H \approx 100 \ \mathrm{Tesla}.\]

\section{Angular dependence of reentrant transition temperature
in Li$_{0.9}$Mo$_6$O$_{17}$}

Our next step is to theoretically explore the variation in
$T^*(\alpha, H)$ due to angular deviation in the optimal
experimental geometry. Depicted in Fig.4 is a magnetic field,
$\vec{H}$, perpendicular to the conducting chains and subtending a
small angle $\alpha$ with respect to the layers. For this
configuration, we take the field and vector potential, along with
the generalization of the Hamiltonian in Eq.(3) to be

\begin{figure}[h]
\centering
\includegraphics[width=95mm, height=65mm]{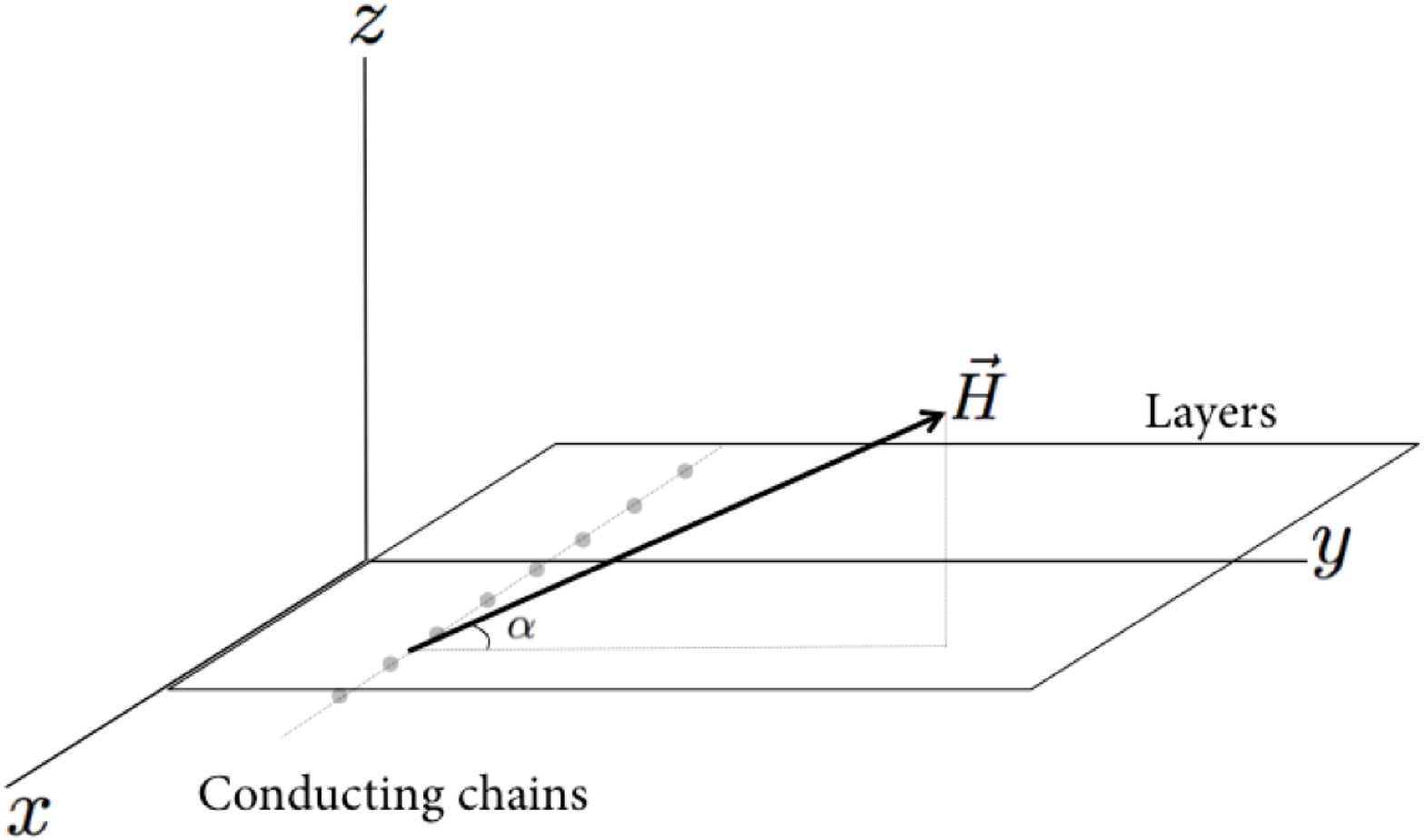}
\caption{Magnetic field is inclined at an angle $\alpha$ with
respect to the layers of a Q1D conductor. The resulting angular
dependence in $T^*(\alpha, H)$ can be used for accuracy in
experimental alignment of Li$_{0.9}$Mo$_6$O$_{17}$ crystals in
magnetic field.}
\end{figure}

\[\vec{H} = (0, H\cos\alpha, H\sin\alpha), \ \ \ \ \vec{A}
= (0, Hx\sin\alpha, -Hx\cos\alpha),\]

\begin{multline}
{\mathcal{\hat H}^{\pm} = \mp i v_F \frac{\mathrm{d}}{\mathrm{d} x}
-2t_z\cos\left(p_za_z + \frac{\omega_z(\alpha)}{v_F}x\right)}\\
 - 2t_y\cos\left(p_ya_y - \frac{\omega_y(\alpha)}{v_F}x\right).
\end{multline}

\noindent By following steps similar to the ones that lead to the
derivation of Eq.(15), we can show that the generalization of the
gap integral equation with a magnetic field inclined at angle
$\alpha$ takes the following form:

\begin{multline}
\Delta(x) = \frac{g}{2} \int_{|x-x'|>d} \frac{2\pi T \Delta(x')
\mathrm{d}x'}{v_F\sinh\left[\frac{2\pi T}{v_F}|x-x'|\right]} \times \\
{J_0\left\{  \frac{8t_y}{\omega_y(\alpha)}
\sin\left[\frac{\omega_y(\alpha)}{2v_F}(x-x')\right]
\sin\left[\frac{\omega_y(\alpha)}{2v_F}(x+x')\right]\right\}\times}\\
{J_0\left\{  \frac{8t_z}{\omega_z(\alpha)}
\sin\left[\frac{\omega_z(\alpha)}{2v_F}(x-x')\right]
\sin\left[\frac{\omega_z(\alpha)}{2v_F}(x+x')\right]\right\}, }
\end{multline}

\noindent where the oscillation frequencies are now defined as

\begin{equation}
\omega_y(\alpha) = \frac{ev_Fa_y}{c}H\sin\alpha, \ \ \ \ \omega_z(\alpha)
= \frac{ev_Fa_z}{c}H\cos\alpha.
\end{equation}

\noindent For this problem we consider Eq.(21) in the high-field reentrant
regime, and introduce a normalized Ginzburg-Landau like trial solution with a
variational parameter $\kappa$:

\begin{equation}
\Delta(x)=(2\kappa/\pi)^{1/4} e^{-\kappa x^2}.
\end{equation}

\noindent With a change of variables $x'-x = (v_F/2\pi T)z$ and the
normalization condition $\int \Delta^2(x)\mathrm{d}x=1$, Eq.(21) can
be recast in the form

\begin{multline}
{\frac{2}{g} =
\int_{-\infty}^{+\infty}\mathrm{d}x\int_{|z|>d_1}\mathrm{d}z
\frac{\Delta(x+z)\Delta(x)}{\sinh(|z|)} \times}\\
{J_0\biggl[ \frac{8t_y}{\omega_y(\alpha)}\sin\left(\frac{\omega_y(\alpha) z}{4\pi T}\right)
\sin\left(\frac{\omega_y(\alpha)(2x+z)}{4\pi T}\right)\biggr] \times}\\
J_0\biggl[
\frac{8t_z}{\omega_z(\alpha)}\sin\left(\frac{\omega_z(\alpha)
z}{4\pi T}\right) \sin\left(\frac{\omega_z(\alpha)(2x+z)}{4\pi
T}\right)\biggr],
\end{multline}

\noindent where $d_1=2\pi T d/v_F$. At high magnetic fields and small angles,
the first Bessel function in Eq.(24) can be expanded to leading term:

\begin{equation}
J_0(\cdot\cdot\cdot)\approx 1 -
\frac{t_y^2\omega_y^2(\alpha)}{4\pi^4T^4}x^2z^2 .
\end{equation}

The trial solution in Eq.(23) along with the expansion in Eq.(25) and Eq.(16)
are substituted into Eq.(24) to evaluate in integrals and optimize the resulting
expression with respect to $\kappa$, giving the maximum transition temperature
as a function of the angle: $T^*(\alpha,H)$. Using the defining condition for
zero-field critical temperature in
Eq.(17), and the value of the integral

\[\int_0^{\infty} \frac{z^2\mathrm{d}z}{\sinh(z)} = \frac{7\zeta(3)}{2},\]

\noindent the resulting expression is

\begin{equation}
\ln\left[ \frac{T^*(H)}{T^*(\alpha,H)}\right] =
\left(\frac{v_F}{2\pi T_c}\right)^2
\frac{7\zeta(3)}{2}\left(\frac{\kappa}{2} +
\frac{\lambda^2(\alpha)}{4\kappa}\right),
\end{equation}

\noindent where $T^*(H) = T^*(\alpha=0, H)$, and

\begin{equation}
\lambda^2(\alpha) = \frac{4t_y^2\omega^2_y(\alpha)}{v^4_F},
\end{equation}

\noindent with $\omega_y(\alpha)$ given in Eq.(22), and $\zeta(3)\approx 1.202$
being the value of the Riemann zeta function. The value of kappa
that minimizes the expression in Eq.(26), thus giving the maximum transition
temperature, $T^*(\alpha,H)$, is

\begin{equation}
\kappa = \frac{\lambda(\alpha)}{\sqrt{2}}.
\end{equation}

\noindent For small angles, $\alpha$, we use the approximation

\begin{equation}
\ln\left[ \frac{T^*(H)}{T^*(\alpha,H)}\right] \approx
\frac{T^*(H) - T^*(\alpha,H)}{T^*(H)},
\end{equation}

\noindent and substitute the result of Eq.(28) into Eq.(26) to
solve for $\omega_y(\alpha)$ and obtain an expression for the
following dependence of $T^*(\alpha, H)$, in the high magnetic
field regime corresponding to the reentrant superconducting phase:

\begin{equation}
T^*(\alpha,H) = T^*(H)\left( 1 - \frac{H}{H_{GL}}\sin\alpha\right),
\end{equation}

\noindent where

\begin{equation}
H_{\mathrm{GL}}  = \frac{4\sqrt{2}\pi^2cT^2_c}{7\zeta(3) t_y
ev_Fa_y}
\end{equation}
is the Ginzburg-Landau upper critical field for $\alpha = 90^0$.
[Note that the obtained angular dependence (30),(31) has a
different meaning than the standard Ginzburg-Landau one since in
our case it is valid only at high magnetic fields.]

The results of Eqs.(30),(31), along with the value of
$T^*(\alpha=0) \simeq 1 \ K$ at $H \simeq 100 \ T$ can be used to
calculate the angular dependence of the reentrant transition
temperature, plotted in Fig.5. We note that the sharp drop
observed in the transition temperature for angles near $\alpha=0$
shows that one needs a careful alignment of a magnetic field
during the corresponding experiment. As follows from Fig.5, the
accuracy of the alignment of the field has to be better than
$\delta \alpha = 0.2^0$ in $({\bf y},{\bf z})$ plane (see Fig.4).
On the other hand, it is known that magnetic fields of order of $H
= 15 \ T$ can destroy superconductivity in
Li$_{0.9}$Mo$_6$O$_{17}$ when the field is aligned parallel to the
conducting chains (see Ref. [19]). Therefore, parallel to the
chains component of the magnetic field has to be less than $\delta
H_{\parallel} \simeq 5 \ T$ in the experiments. Thus, accurate
angular orientation is important for detection of the reentrant
superconducting phase at fields of order of $H \simeq 100 \ T$,
where small inclination of the field (in particular, towards the
${\bf z}$ axis) can destroy superconductivity.

\begin{figure}
\centering
\includegraphics[width=80mm, height=65mm]{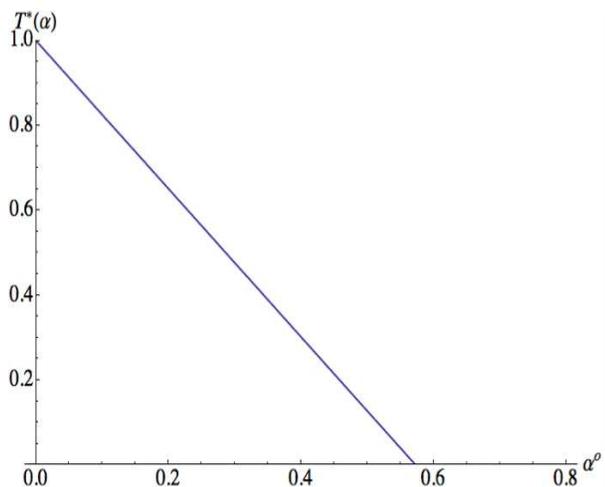}
\caption{Angular variation of reentrant superconducting transition
temperature (in Kelvin), $T^*(\alpha)$, at $H=100 \ T$ for small inclination
angles, $\alpha$ (in degrees) from the optimal experimental
geometry. }
\end{figure}

\section{Conclusion}

In this paper, we have studied the quantum limit reentrant
superconductivity phenomenon in the layered Q1D conductor
Li$_{0.9}$Mo$_6$O$_{17}$. Our results show that superconductivity
can be restored and potentially experimentally detected in this
compound at the reentrant transition temperature $T^*(H=100 \ T)
\simeq 1 \ K$ when a field of order $H \simeq 100$ Tesla is
aligned parallel to the layers and perpendicular to the conducting
chains. We noted that such magnetic fields are currently
experimentally available as pulsed non-destructive fields.
Furthermore, we have specified how the reentrance transition
temperature, $T^*(\alpha, H)$ varies for arbitrary, as well as
small angular deviations from the optimal experimental geometry.
This information is important for accurate alignment of a sample
in magnetic field. If confirmed experimentally, the reentrant
superconducting phase in Li$_{0.9}$Mo$_6$O$_{17}$ would be the
first example of survival of superconductivity in ultra-high
magnetic fields and would in addition unequivocally confirm
spin-triplet pairing nature in this compound. $\\$ $^*$ Also, at
L.D. Landau Institute for Theoretical Physics, RAS, 2 Kosygina
Street, Moscow 117334, Russia.

\end{document}